\documentclass[11pt]{article}
\usepackage{amssymb}
\usepackage{float}
\usepackage{times}
\usepackage{graphicx}
\usepackage{amsmath}
\usepackage{wrapfig}
\def\Znote#1{}

\def\midformat{
\setlength{\textheight}{8.9in}
\setlength{\textwidth}{6.7in}
\setlength{\evensidemargin}{-0.2in}
\setlength{\oddsidemargin}{-0.2in}
\setlength{\headheight}{0in}
\setlength{\headsep}{10pt}
\setlength{\topsep}{0in}
\setlength{\topmargin}{0.0in}
\setlength{\itemsep}{0in}       
\renewcommand{\baselinestretch}{1.1}
\parskip=0.070in
}

\newtheorem{theorem}{Theorem}[section]

\newtheorem{definition}{Definition}
\newtheorem{claim}{Claim}
\newtheorem{lemma}[theorem]{Lemma}

\newtheorem{corollary}[theorem]{Corollary}

\newtheorem{comment}{Comment}

\def\boldhead#1:{\par\vskip 7pt\noindent{\bf #1:}\hskip 10pt}
\def\ithead#1:{\par\vskip 7pt\noindent{\it #1:}\hskip 10pt}
\def\inline#1:{\par\vskip 7pt\noindent{\bf #1:}\hskip 10pt}
\long\def\comment #1\commentend{}
\long\def\commfull #1\commend{#1}
\long\def\commabs #1\commenda{}
\long\def\commtim #1\commendt{#1}
\long\def\commb #1\commbend{}
\def\blackslug{\hbox{\hskip 1pt \vrule width 4pt height 8pt
    depth 1.5pt \hskip 1pt}}
\def\QED{\hfill \ensuremath{\square}}

\def\inQED{~~~~~\quad\blackslug\lower 8.5pt\null}
\def\Proof{\noindent{\bf Proof:~}}
\def\ProofOf#1#2{\noindent{\bf Proof of #1 \ref{#2}:~}}

\long\def\PPP#1{\noindent{\bf Proof:}{ #1}{\quad\blackslug\lower 8.5pt\null}}

\long\def\denspar #1\densend
{#1}
\def\DEF{\stackrel{\rm def}{=}}
\newcommand{\Eqr}[1]{Eq.~(\ref{#1})}

\newcommand{\OO}[1]{O\left( #1\right)}
\newcommand{\OM}[1]{\Omega\left( #1 \right)}
\newcommand{\Set}[1]{\left\{ #1 \right\}}

\newcommand{\ignore}[1]{}

\def\eps{\epsilon}

\newif\ifnotesw\noteswtrue
   {\ifnotesw\marginpar[\hfill\(\top\)]{\(\top\)}\fi}%
   {\ifnotesw\marginpar[\hfill\(\bot\)]{\(\bot\)}\fi}

\newcommand{\mnote}[1]%
    {\ifnotesw\marginpar%
        [{\scriptsize\begin{minipage}[t]{\marginparwidth}
        \raggedleft#1%
                        \end{minipage}}]%
        {\scriptsize\begin{minipage}[t]{\marginparwidth}
        \raggedright#1%
                        \end{minipage}}%
    \fi}

\newcounter{tempenumi}
\newenvironment{denselist}{\setcounter{tempenumi}{\value{enumi}}
\begin{list}{(\arabic{enumi})}{
\usecounter{enumi}
\setlength{\topsep}{0pt} \setlength{\partopsep}{0pt} \setlength{\itemsep}{0pt}
}}{\end{list}\setcounter{enumi}{\value{tempenumi}}}

\newenvironment{subdenselist}{
\begin{list}{(\arabic{enumi}\alph{enumii})}{
\usecounter{enumii}
\setlength{\topsep}{0pt}
\setlength{\partopsep}{0pt}
\setlength{\itemsep}{0pt}
}}{\end{list}}

\newcommand{\headersmall}[3]{
   \pagestyle{plain}
   \noindent
      \vbox{
    \hbox to 6.28in { \Course \hfill #2}
\hrule

\vspace{1mm}

Prof. Patt-Shamir

       \vspace{7mm}

       \hbox to 6.28in {\Large\bf \hfill #3 \hfill}
      }
   \vspace*{8mm}
}

\newcommand{\tauheadersmall}[4]{
   \pagestyle{plain}
   \noindent
      \vbox{
   \vspace{-.3in}
    \hbox to 6.5in {\sf #1 \hfill #3}
\hrule
#2\hfill
       \vspace{7mm}
       \hbox to 6.28in {\Large\bf \hfill #4 \hfill}
      }
   \vspace*{-6mm}
}

\newcommand{\lectureheader}[3]{
   \pagestyle{plain}
   \noindent
      \vbox{
    \hbox to 6.28in { \Course
                        \hfill #2 }
\hrule
\vspace*{1mm}
Lecturer: #3
       \vspace{7mm}\\
       \hbox to 6.28in {\LARGE\bf \hfill Lecture #1 \hfill}
        }
   \vspace*{4mm}
}

\newcommand{\newlectureheader}[3]{
   \pagestyle{plain}
   \noindent
      \vbox{
    \hbox to 6.28in { 6.04s Design and Analysis of Distributed Protocols
                       \hfill #2}
\hrule
\vspace*{1mm}
Lecturer: #3
       \vspace{7mm}\\
       \hbox to 6.28in {\LARGE\bf \hfill Lecture #1 \hfill}
        }
   \vspace*{4mm}
}

\newcommand{\talkheader}[3]{
   \pagestyle{plain}
   \vbox{
    \hbox to 6.28in {Lecturer: #3 \hfill #2 }
    \hrule
    \vspace*{7mm}
    \hbox to 6.28in {\LARGE\bf \hfill #1 \hfill}
   }
   \vspace*{4mm}
}

\def\Plus{\hbox{\raise 0.3ex\hbox{\tiny +}}} 

\def\mtoday{\ifcase\month\or
  January\or February\or March\or April\or May\or June\or July\or
  August\or September\or October\or November\or
  December\fi\space\number\year}

\floatstyle{ruled}
\newfloat{algorithm}{thp}{lop}
\floatname{algorithm}{Algorithm}

\newcommand{\abs}[1]{\left\vert {#1} \right\vert}
\newcommand{\inv}[1]{\frac{1}{{#1}}}
\newenvironment{denseenumerate}{
\begin{list}{\arabic{enumi}.}{
\usecounter{enumi}
\setlength{\topsep}{0pt} \setlength{\partopsep}{0pt} \setlength{\itemsep}{0pt}
}}{\end{list}}
\def\bbE{{\mathbb E}}

\def\congest{{\mathbf{CONGEST}}}
\def\dnc{\mathsf{DistNearClique}}
\def\poly{{\rm poly}}

\renewcommand{\mid}{\;:\;}
\midformat

\renewcommand{\paragraph}[1]{\par\noindent\textbf{#1}}
\newcommand{\ENC}[1]{#1-near clique}
\newcommand{\ENCe}{$\eps$-near clique}

\title{\textbf{Distributed Discovery of Large Near-Cliques}}
\author{
Zvika Brakerski\\
Dept.\ of Computer Science and Applied Mathematics\\
Weizmann Institute of Science\\
Rehovot 76100\\
Israel\\
{\small\tt zvika.brakerski@weizmann.ac.il}
\and Boaz Patt-Shamir\thanks{
Supported in part by the Israel Science Foundation, grant 664/05.
}\\
Dept.\ of Electrical Engineering\\
Tel Aviv University\\
Tel Aviv 69978\\
Israel\\
{\small\tt boaz@eng.tau.ac.il} }

\begin{document}
\def\thepage{}
\begin{titlepage}
\maketitle
\begin{abstract}
  Given an undirected graph and $0\le\eps\le1$, a set of nodes is
  called \ENC{$\eps$}\ if
  all but an $\epsilon$ fraction of the pairs of nodes in the set have
  a link between them. In this paper we present a fast synchronous
  network algorithm that uses small messages and finds a near-clique.
  Specifically, we present a constant-time 
  algorithm that finds, with constant probability of success, a linear size
  \ENC{$\eps$} if there exists an \ENC{$\eps^3$} of linear size
  in the graph.  The algorithm uses messages of $O(\log n)$ bits. The
  failure probability can be reduced to $n^{-\Omega(1)}$ in $O(\log
  n)$ time, and the algorithm also works if the graph contains a clique of
  size $\Omega(n/\log^{\alpha}\log n)$ for some $\alpha \in (0,1)$.
  \Znote{Corrected (added $\alpha$ to the original $\OM{n/\log\log n}$).}
  Our approach is based on a new idea of
  adapting property testing algorithms to the
  distributed setting. 
\end{abstract}

\Znote{Some more things I don't like: theorem/corollary number by section - not
needed for such a small submission. Theorem 2.1 and 4.7 cannot be two separate
theorems, it doesn't make sense.}

\end{titlepage}

\pagenumbering{arabic}

\section{Introduction}
Discovering dense subgraphs is an important task both theoretically
and practically.  From the theoretical point of view, clique detection
is a fundamental problem in the theory of computational complexity,
and for distributed algorithms,
computing useful constructs of the underlying communication graph is
one of the central goals. Let us elaborate a little about that.

Dense graph detection has always been an important problem for
clustering and hierarchical decomposition of large systems for
administrative purposes, for routing and possibly other
purposes \cite{BasangiMPP-06}. Another reason to consider dense
subgraphs is conflicts in radio ad-hoc networks \cite{GuptaW04}.
On top of these low-level communication-related tasks, dense subgraph
detection has recently also attracted
considerable interest for Web analysis:
as is well known, the ranking of results
generated by search engines such as Google's PageRank \cite{pagerank}
is derived from the topology of the Web graph; in particular, it can
be heavily influenced by ``tightly knit communities'' \cite{salsa},
which are essentially dense subgraphs. Hence, to understand the structure
of the web, it is important to be able to
identify such communities. Another dimension where dense
subgraphs are interesting for the Web is time: it has been observed
\cite{KumarNRT-05} that evolution of links in blogs is, to some
extent, a sequence of significant events, where significant events are
characterized as dense subgraphs. Thus, considering the web as a
dynamic graph, identifying large dense subgraphs is useful in
understanding its temporal aspect.

\textbf{Our Contribution.}
 In this paper we give an efficient randomized distributed algorithm that finds
large dense subgraphs.
Obviously, our algorithm does not decide whether there exists a large clique
in the graph: that would be impossible to do efficiently unless P=NP.
Instead, our algorithm solves a
relaxed problem.  First, 
we find near-cliques, defined as follows. Given a
graph and a constant $\epsilon>0$, a set of nodes $D$ is said to be
an \emph{\ENC{$\eps$}} if all, except perhaps an $\eps$ fraction of the
pairs of nodes of $D$ have an edge between them (see Section
\ref{sec:model} for more details). For example, using this definition,
a clique is
\ENC{$0$}. Second, our algorithm only identifies a large near-clique,
and it is only guaranteed that the density of the output is close to
the best possible. For example, given a
graph $G$ and a constant $\epsilon>0$ such that $G$ contains an
\ENC{$\eps$} with a linear number of nodes, our algorithm finds at
least one \ENC{$\eps^{1/3}$} of linear size in $G$. (Our algorithm
can also discover dense subgraphs of sublinear size for smaller values
of $\eps$.)
Our algorithm is extremely frugal: the output is computed (with
constant probability of success) in constant number of rounds,
and all messages contain $O(\log n)$ bits.%
\footnote{
If messages may be of unbounded size, the problem becomes both trivial
(from the communication viewpoint) and infeasible (from the
computation viewpoint). See Section \ref{sec-simple}.
}
Given any $q>0$, it is possible to amplify the success
probability to
$1-q$  in  $O(\log(1/q))$ time.

In addition to the direct contribution of the algorithm, we believe
that our methodology is interesting in its own right. Specifically,
our work extends ideas presented in \cite{GGR} in relation to property
testing of the $\rho$-clique problem (defined below). Even though our
construction does not use the property tester of \cite{GGR} as a black
box, our approach of deriving a distributed algorithm from graph
property testers seems to be an interesting idea to consider when
approaching  other problems as
well. In a nutshell,  property testers do very
little overall work but have a ``random access'' probing capability, namely
they can probe topologically distant edges;  distributed
algorithms, on the other hand, can do a lot of work (in parallel), but
information flow is local, i.e.,
an algorithm which runs for $T$ rounds allows each node to gather
information only from distance at most $T$. However, quite a few
graph property testers exhibit some locality that can be exploited by
distributed algorithms.

\textbf{ Related work.}
We are not aware of any previous distributed algorithm that finds large
dense subgraphs efficiently.
Maximal independent sets, which are cliques in the
complement graph, can be found efficiently distributively
\cite{L-85,AlonBI86}.
In this case, there can be no non-trivial guarantee about their size with
respect to the size of the largest (maximum) independent set in the
graph. But on the positive side, the sets output by these algorithms are
strictly independent.

Much more is known about dense subgraphs in the centralized setting.
The fundamental result is that finding the largest \emph{clique} (i.e., fully
connected subset of nodes) in a graph, or even approximating its size
to within a factor of $n^{1-\epsilon}$ for any constant $\epsilon>0$,
is computationally hard \cite{Hastad-clique}.
There are some closely related results in the centralized model and in the
property testing model.  In the centralized model, the Dense $k$-Subgraph (DkS)
problem was studied. In DkS, the input consists of a graph and a positive
integer $k$, and the goal is to find a the subset of $k$ nodes with the most
number of edges between them. Feige, Peleg and Kortsarz \cite{FeigeKP-01}
present a centralized algorithm approximating DkS within a factor of
$O(n^\delta)$ for a certain $\delta<1/3$, and it is also possible to approximate DkS
to within roughly $n/k$ \cite{FeigeL-01}. Abello, Resende and Sudarsky
\cite{QuasiCliques} presented a heuristic for finding near-cliques (which they
refer to as ``Quasi-Cliques'') in sparse graphs.

%

Property testing was defined by Rubinfeld and Sudan \cite{RS96} for algebraic
properties, and extended by Goldreich, Goldwasser and Ron \cite{GGR} to
combinatorial graph properties.  The relevant concepts are the following. In
the \emph{dense graph model}, the basic action of a property tester is to query
whether a pair of nodes is connected by an edge in the graph.  An $n$-node
graph is said to have the $\rho$-clique property if it contains a clique of
size $\rho n$, for some given parameter $0\le \rho\le 1$. The $\rho$-clique
tester of \cite{GGR} gets an $n$-node graph $G$ and constants $\rho, \epsilon$
as input, and decides, using $\tilde O\left(1/\epsilon^6\right)$ queries and
with constant probability of being correct, whether the input graph has a
$\rho$-clique or whether no set of $\rho n$ nodes in $G$ is
\ENC{$(\eps/\rho^2)$}. \Znote{It decides whether the graph is $\eps$-far from
having a $\rho$-clique which translates into \ENC{$(\eps/\rho^2)$}.}
\ignore{There is no guarantee regarding the answer when the input graph
contains an \ENC{$\eps$} of size $\rho n$.} \Znote{Removed sentence here that
is wrong (as the next paragraph shows).} They further present an ``{approximate
find}'' algorithm that, provided that the property tester answers in the
affirmative, finds an \ENC{$\eps$} of size $\rho n$ in the graph in $O(n)$
time. Our algorithm is a new variant of the ideas of \cite{GGR} and, using a
new analysis, gets a better complexity result in the case of the relaxed
assumption of existence  of a near-clique.

This relaxation is a special case of \emph{tolerant} property testing
\cite{PRR06}, which in our case can be defined as follows. An
$(\epsilon_1, \epsilon_2)$-tolerant $\rho$-clique tester takes
parameters $\rho, \epsilon_1$ and $\epsilon_2$ where $\eps_1<\eps_2$,
and decides whether the graph contains an \ENC{$\epsilon_1$} or
whether no set of $\rho n$ nodes is an \ENC{$\epsilon_2$}. The general
results of \cite{PRR06} imply that the property tester of \cite{GGR} is in fact
$(\epsilon^6, \epsilon)$-tolerant (our construction is $(\epsilon^3,
\epsilon)$-tolerant).
Fischer and Newman \cite{FN05} prove
a general result (for any property testable in $O(1)$ queries), whose
implication to our case is that it is possible to find the smallest
$\epsilon$ for which a graph has an \ENC{$\epsilon$} of size $\rho n$,
but the  query complexity is an exponent-tower of height poly($\epsilon^{-1}$).

A relation between distributed algorithms and property testers was pointed out
by Parnas and Ron in \cite{PR-07}, where it is shown for Vertex Cover how to
derive a good property tester from a good distributed algorithm (the reduction
goes in the direction opposite to the one we propose in this paper). Recently,
techniques from property testing were  used, along with other techniques by
Nguyen and Onak \cite{NO08},  to present constant-time approximation algorithms
for vertex-cover and maximum-matching in bounded-degree graphs. Their
techniques also yield constant-time distributed algorithms for these problems.
Saks and Seshadhri \cite{SaksS08} show how to devise a parallel algorithm that
``reconstructs'' a noisy monotone function, again using ideas from property
testing.

\textbf{Paper organization.} The problem and
main results are stated in
Section \ref{sec:model}. Simple solutions are discussed in Section
\ref{sec-simple}. The algorithm is presented  in Section
\ref{sec:alg} and analyzed in Section \ref{sec-analysis}.
We conclude in  Section
\ref{sec:discussion}. Some proofs are presented in an appendix.


\section{Definitions, Model, Results}
\label{sec:model}
\paragraph{Graph concepts.}
In this paper we assume that we are given a simple undirected graph
$G=(V,E)$.  We denote $n\DEF|V|$.
For any given set $U\subseteq V$ of nodes, $\Gamma(U)$ denotes the set of all
  neighbors of nodes in $U$. Formally,
$\Gamma(U)\DEF\Set{v\mid u\in U\mbox{ and } (u,v)\in E}$.

For counting purposes, we use a slightly unusual approach, and view each
undirected edge $\Set{u,v}$ as two anti-symmetrical directed edges
$(u,v)$ and $(v,u)$. Using this approach, we define the following
central concept.
\begin{definition}
  Let $G=(V,E)$ be a graph. A set of nodes $D\subseteq V$ is called
  \emph{\ENC{$\eps$}} if
$$
\Big|\Set{(u,v)\mid (u,v)\in D\times D\mbox{ and }\Set{u,v}\in E}\Big|
~~\ge~~(1-\eps)\cdot|D|\cdot(|D|-1).
$$
\end{definition}
In such case we also say that the \emph{density} of $D$ is at least $1-\eps$.
\paragraph{Distributed Algorithms.}
We use the standard synchronous distributed model $\congest$ as defined in
\cite{Peleg:book}.  Briefly, the system is modeled by an undirected graph,
where nodes represent processors and edges  represent communication links. It
is assumed that each node has a unique $O(\log n)$ bit identifier. An execution
starts synchronously and proceeds in rounds: in each round each node sends
messages (possibly different messages to different neighbors), receives
messages, and does some local computation. By the end of the execution, each
processor writes its output in a local register.  A key constraint in the
$\congest$ model is that the messages contain $O(\log n)$ bits, which
intuitively means that each message can describe a constant number of nodes,
edges, and polynomially-bounded numbers. The time complexity of the algorithm
is the maximal number of rounds required to compute all output values. We note
that we assume no processor crashes, and therefore any synchronous algorithm
can be executed in an asynchronous environment using a synchronizer
\cite{A-85}.

\paragraph{Problem Statement.}
In this paper we consider algorithms for finding \ENCe.
The input to the algorithm is the underlying
communication graph and $\epsilon$. Each node has an output register,
which holds, when the
algorithm terminates, either a special value ``$\bot$'' or a
label. All nodes with the same output label are in the same
\ENCe, and $\bot$ means that the node is not associated with
any near-clique. Note that there may be more than one near-cliques in
the output.

\paragraph{Results.}
The main result of this paper is given below (see Theorem
\ref{thm-detailed} for a detailed version).

\newcounter{algthm}\setcounter{algthm}{\value{theorem}}
\begin{theorem}\label{thm:algorithm-hl}
  Let $\eps,\delta >0$. If there exists an \ENC{$\epsilon^3$} $D
  \subseteq V$ with $\abs{D}\ge\delta n$, then an
  \ENC{$O(\eps/\delta)$} $D'$ with $|D'|=|D|\cdot(1-O(\eps))$ can be found
  by a distributed algorithm with probability $\Omega(1)$, in
  $2^{\OO{\eps^{-4}\delta^{-1}\log (\eps^{-1}\delta^{-1})}}$
  rounds, using messages of $O(\log n)$ bits.
\end{theorem}
We stress that the message length is a function of $n$ and  is independent of
$\epsilon, \delta$.

Let us list a few immediate corollaries to our result. First, for the
case where there are near-cliques
of linear size (i.e., $\delta=\OM{1}$).

\begin{corollary}
  Let $\eps$ be a constant. If there exists an \ENC{$\epsilon^3$} $D
  \subseteq V$ with $\abs{D}=\Theta(n)$, then an \ENC{$O(\eps)$} $D'$
  with $|D'|=|D|\cdot(1-O(\eps))$ can be found by a distributed algorithm
  with probability $\Omega(1)$, in $O(1)$ rounds and using messages of
  $O(\log n)$ bits.
\end{corollary}

Second, for the
case where there are strict cliques of (slightly) sublinear size.
\begin{corollary}
  If there exists a clique $D$ with $\abs{D}\ge n/\log^{\alpha}\log n$ for
  a sufficiently small constant $\alpha>0$, then an \ENC{$o(1)$} $D'$ with
  $|D'| \ge (1-o(1))\cdot\abs{D}$ can be found by a distributed
  algorithm with probability $1-o(1)$, in polylogarithmic number of
  rounds and using messages of $O(\log n)$ bits.
\end{corollary}

\section{Simple Approaches}
\label{sec-simple}
In this section we consider, as a warm-up, two simplistic approaches
to solving the near-clique problem, and explain why they fail.

\textbf{The neighbors' neighbors algorithm.}
The first idea is to let each node inform all its neighbors about all
its neighbors. This way, after one communication round, each node
knows the topology of the graph to distance 2, and can therefore find
the largest clique it is a member of. It is easy to kill cliques that
intersect larger cliques (using, say, the smallest ID of a clique as a
tie-breaker), and so we can output a set of locally largest cliques in
a constant number of rounds. Indeed, one can develop a correct
algorithm based on these ideas, but there are two show-stopper
problems in this case. First, the size of a message sent in this
algorithm may be very large: a message may contain all node IDs. (This
is the \textbf{LOCAL} model \cite{Peleg:book}.).
And second, the algorithm requires each node to locally solve the
largest clique problem, which is notoriously hard to compute. We thus
rule out this algorithm on the basis of prohibitive computational and
communication complexity.

\textbf{The shingles approach.}
Based on the idea of shingles \cite{BroderGMZ-97}, one may consider
the following algorithm. Each node picks a random ID (from a space
large enough so that the probability of collision is negligible),
sends it out to all its neighbors, and then selects the smallest ID it
knows (among its neighbors and itself) to be its label. All nodes with
the same label are said to be in the same \emph{candidate set}. Each
candidate set finds its density by letting all nodes send their degree
in the set to the set leader (the namesake of the set label), and only
sets with sufficient size and density survive. Conflicts due to
overlapping sets are
resolved in favor of the larger set, and if equal in size, in favor of
the smaller label. Call this the ``shingles algorithm.''

Clearly, if there is a clique of linear size in
the graph, then with probability $\Omega(1)$  the globally minimal ID
will be selected by a node in the clique, in which case all nodes in
the clique belong to the same candidate set. Unfortunately, many other
nodes not in the clique may also be included in that candidate set,
``diluting'' it significantly.
Formally,  we claim the following.
\begin{claim}
\label{cl-shingles}
For any constant $\delta \in (0,1)$ there exists an infinite family of
graphs $\{G_n\}$ such that $G_n$ has $n$ nodes and it contains a
clique of size $\delta n$, but for all $\epsilon <
\min\left\{\frac{1-\delta}{1+\delta}, 1/9\right\}$ and for
sufficiently large $n$, the shingles algorithm cannot
find an \ENC{$\epsilon$} with at least
$(1-\epsilon)\delta n$ nodes in $G_n$.
\end{claim}

\begin{wrapfigure}{r}{1.9in}
  \centering
  \includegraphics[width=1.5in]{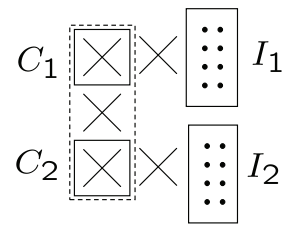}
  \caption{\em\small Crosses represent full connectivity.}
  \label{fig-shingles}
\end{wrapfigure}

\Proof
Fix $\delta\in (0,1)$ and consider, for simplicity, $n$ such that
$\delta n$, $n$ are even. The graph $G_n$ is defined as follows. The
nodes of $G_n$ are partitioned into
four sets denoted $C_1, C_2, I_1, I_2$, where
$\abs{C_1}=\abs{C_2}=\delta n/2$, $\abs{I_1}=\abs{I_2}=(1-\delta)n/2$.
The sets $C_1, C_2$ are complete subgraphs and $I_1, I_2$ are
independent sets (see Figure \ref{fig-shingles}).  The pairs of sets
$(I_1, C_1)$, $(C_1, C_2)$, $(C_2, I_2)$ are connected with complete
bipartite graphs (i.e., every node in $I_1$ is connected to every node
in $C_1$ and similarly for the other pairs). The resulting graph
contains a clique $C=C_1\cup C_2$ of size $\delta n$.

We proceed by case analysis. Let $v_{\min}$ denote the node
with the globally minimal ID in $G_n$, as drawn by the shingle
algorithm.
\\
\emph{ Case 1: $v_{\min} \in C_1 \cup C_2$.} W.l.o.g assume that $v_{\min} \in C_1$. Then
$v_{\min}$'s candidate set contains exactly $C_1 \cup C_2 \cup I_1$, a set whose
density is
\[
\frac{\binom{\abs{C_1}+\abs{C_2}}{2}+\abs{I_1}\cdot\abs{C_1}}{\binom{\abs{C_1}+\abs{C_2}+\abs{I_1}}{2}}
= \frac{\binom{\delta n}{2}+\delta(1-\delta)n^2/4}{\binom{(1+\delta)n/2}{2}} =
\frac{2\delta}{1+\delta}~,
\]
and for $\epsilon < \frac{1-\delta}{1+\delta}$ the density is less
than $1-\epsilon$. Clearly in this case all other candidates are
subsets of $I_1 \cup I_2$ and thus have density $0$.
\\
\emph{Case 2: $v_{\min} \in I_1 \cup I_2$.} W.l.o.g assume that
$v_{\min} \in I_1$.  Then $v_{\min}$'s candidate set is exactly $C_1 \cup
\{v_{\min}\}$ and thus has size $\delta n/2 + 1$ which is asymptotically
smaller than $(1-\epsilon) \delta n$ for any constant $\epsilon <1/2$.
\\
Finally, consider the other candidate sets in this case. Clearly all nodes
in $C_2$ belong to the same candidate set. Let $A$ denote the set of
vertices from $I_1 \cup I_2$ belonging to $C_2$'s candidate set. If
$\abs{A} < \delta n /4$ then the candidate set size is $\abs{C_2}+\abs{A}
< 3\delta n /4$ which is less than $(1-\epsilon)\delta n$ for all
$\epsilon \le 1/4$. If $\abs{A} \ge \delta n/4$ then the candidate set
density is at most
\[
\frac{\binom{\abs{C_2}}{2}+\abs{C_2}\cdot\abs{A}}{\binom{\abs{C_2}+\abs{A}}{2}}
\le 1-\frac{1-4/\delta n}{3\cdot(3-4/\delta n)}
\]
which is asymptotically less than $1-\epsilon$ for any $\epsilon$ smaller than
$1/9$.  The remaining candidate sets are subsets of $I_1 \cup I_2$ and thus
have density $0$.
\QED

\textbf{Summary.}
The simple approaches demonstrate the basic difficulty of the distributed
\ENC{$\epsilon$} problem: looking to distance 1 is not sufficient, but
looking to distance 2 is too costly. The algorithm presented next
finds a middle ground using sampling.

\section{Algorithm}
\label{sec:alg}

Below we present the algorithm for finding dense subgraphs.
Analysis is presented in Section \ref{sec-analysis}.



\paragraph{The basic idea.}
Let $V'\subseteq V$ be a set of nodes. Define $K(V')$ to be the set of
all nodes which are adjacent to {all} other nodes in $V'$, i.e.,
$K(V')\DEF\Set{v\mid \Gamma(v)\supseteq V'\setminus\Set{v}}$.  Further
define $T(V')$ to be the set of nodes in $K(V')$ that are adjacent to
all nodes in $K(V')$, i.e., $T(V')\DEF\Set{v\in K(V')\mid
  \Gamma(v)\supseteq K(V')\setminus\Set{v}}$.  Our starting point is
the following key observation (essentially made in \cite{GGR}). If $D$
is a clique, then $D\subseteq K(D)$, and also, by definition,
$D\subseteq T(D)$.  Furthermore, $T(D)$ is a clique since each $v \in
T(D)$ is adjacent to all vertices in $K(D)$ and in particular those in
$T(D)$.

 The algorithm finds a set which is roughly $T(D)$, where $D$ is
the existing near-clique, by random sampling. Suppose
that we are somehow given a random sample $X$ of $D$. Consider $K(X)$:
it is possible that $K(X)\not\subseteq K(D)$, because $K(X)$ is the set
of nodes that are adjacent to all nodes in $X$, but not
necessarily to all nodes in $D$. We therefore relax the definitions of
$K(X)$ and $T(X)$ to approximate ones $K_\eps(X)$ and $T_\eps(X)$.
Finally, we overcome the difficulty of
inability to sample $D$ directly (because $D$ is unknown), by taking a
random sample $S$ of $V$,
trying \emph{all} its subsets $X\subseteq S$ ($|S|$ is polynomial in
$1/\eps$), and outputting the
maximal $T(X)$ found.



\paragraph{Description and implementation details.}
We now present the algorithm in detail.
We shall use the following notation. Let  $X\subseteq V$
be a set of nodes, and let $0\le \eps\le 1$. We denote by $K_\eps(X)$ the set
of nodes which are neighbors of all but an $\eps$-fraction of the nodes in $X$,
i.e.,
\begin{equation}
K_\eps(X)~\DEF~\Set{v\in V\mid |\Gamma(v)\cap X|\ge(1-\eps)|X|}~.
\end{equation}
Using the notion of $K_\eps$, we also define
\begin{equation}
T_\eps(X)~\DEF~K_\eps(K_{2\eps^2}(X))\,\cap\, K_{2\eps^2}(X)~.
\end{equation}


\newpage

\def\Par{\textsf{parent}}
\def\Root{\textsf{root}}
\def\Comp{\textsf{Comp}}
\def\NULL{\texttt{NULL}}
\def\lab{\textsf{label}}

{\hrule height.8pt depth0pt \kern2pt}
\noindent\textbf{Algorithm $\dnc$}\par
{\kern2pt\hrule\kern2pt} \vspace{1mm}
{\small
\paragraph{Input}: Graph $G=(V,E)$, $\epsilon > 0$, $p \in (0,1)$.
\paragraph{Output}:
A label $\lab_v\in V\cup\Set{\bot}$ at each node $v$, such that $u$ and $v$ are in the
same near clique iff $\lab_v=\lab_u\ne\bot$.
\par
{\kern2pt\hrule\kern2pt} \vspace{1mm}

\paragraph{Sampling stage.}
Each node joins a set $S$ with probability $p$ (i.i.d). Let $G[S]$ denote the
subgraph of $G$ induced by $S$.

\paragraph{Exploration stage:} Finding near-clique ``candidates''.
\begin{denselist}
\item\label{st-B-tree}
  Construct a rooted spanning tree for  each connected component
  of $G[S]$. By the end of this step, each node  $v\in S$ has a
  variable $\Par(v)$ that points to one of its neighbors (for the root,
  $\Par(v)=\NULL$).
\item\label{st-B-ids}
  Each node in $S$ finds the identity of all nodes in its
  connected component and stores them in a variable  $\Comp(v)$.
\item\label{st-B-neigh}
  Each node $v\in S$  sends $\Comp(v)$ to all its
  neighbors $\Gamma(v)$. A node $u\in\Gamma(S)$ may receive at this step
  messages from several nodes, that may or may not be in different
  components of
  $S$.
  Each node $u\in\Gamma(S)$ sets a parent pointer
  $\Par_{S_i}(u)$ for each connected component $S_i$ of $G[S]$ that $u$ is
  adjacent to (choosing arbitrarily between its neighbors from the
  same $S_i$).
\item\label{st-B-count}
  Let $u\in\Gamma(S)$. Let $S_1,\ldots,S_\ell$
  be the different connected components which are adjacent to $u$. For
  each $S_i$ where $1\le i\le\ell$, the following procedure is
  executed.
  \begin{subdenselist}
  \item \label{st-B-det}
    For \emph{all} subsets $X\subseteq S_i$, $u$ determines (using the
    information received in Step \ref{st-B-neigh}) if
    $u\in K_{2\eps^2}(X)$.
  \item \label{st-B-add}
    $u$ sends the results of the computations
    ($2^{|S_i|}$ bits) to all its neighbors, including
    $\Par_{S_i}(u)$.
  \item \label{st-B-convg}
    This information is sent up to
    the root of $S_i$, summing the counts for each $X$ along the way, so
    that the root of $S_i$ knows the value of $|K_{2\eps^2}(X)|$ for
    each $X\subseteq S_i$.
  \item \label{st-B-bcast}
    The root sends the value of  $|K_{2\eps^2}(X)|$ down back to
    all nodes in $\Gamma(S_i)$.
  \item \label{st-B-K2} Each node $v\in K_{2\eps^2}(X)$ sends
    $|K_{2\eps^2}(X)|$ to all its neighbors, for each $X\subseteq
    S_i$.
  \item \label{st-B-dec}
    Each node $u\in \Gamma(S_i)$
    finds whether $u\in K_\eps(K_{2\eps^2}(X))$ for each $X\subseteq
    S_i$, and thus determines whether $u\in T_\eps(X)$ for each $X$.
  \end{subdenselist}
\end{denselist}

\paragraph{Decision stage:} Conflict resolution.
\begin{denselist}
  \item\label{st-C-size} For each connected component $S_i$, the size of $T_\eps(X)$ is
    computed for each $X\subseteq S_i$ similarly to Steps
    \ref{st-B-add}--\ref{st-B-convg} of the exploration stage. Let $X(S_i)$ be the
  subset that maximizes $|T_\eps(X)|$ over all $X\subseteq S_i$.
  \item The root of each component $S_i$ sends $|T_\eps(X(S_i))|$ out to all
  nodes in $\Gamma(S_i)$.
  \item\label{st-C-abort} After receiving $|T_\eps(X(S_i))|$ for all relevant connected components,
  each node sends an ``acknowledge'' message to the component reporting the largest
  $|T_\eps(X(S_i))|$, breaking ties in favor of the largest root ID, and an
  ``abort'' message to all other components.
  \item If no node in $\Gamma(S_i)$ sent an ``abort'' message to $S_i$, the root
  sends back the result to all nodes in $T_\eps(X(S_i))$ (this is done
  by sending $X(S_i)$). The label of a node in $T_\eps(X(S_i))$ is
  the root ID of $S_i$, and $\bot$ otherwise.
\end{denselist}
}  {\kern2pt\hrule\relax}

\vspace*{2mm}

\Znote{Added notation for $G[S]$ because ``connected component of $S$'' is a
confusing terminology.} The algorithm works in stages as follows. In the
\textbf{sampling stage}, a random sample of nodes $S$ is selected; the
\textbf{exploration stage} generates near-clique candidates by considering
$T_{\eps}(X)$ for all $X \subseteq S_i$ s.t.\ $S_i$ is a connected component of
the induced subgraph $G[S]$; and the \textbf{decision stage} resolves conflicts
between intersecting candidates. Pseudo-code for Algorithm $\dnc$ is presented
above. A detailed explanation of the distributed
implementation of Algorithm $\dnc$ follows.

The \textbf{sampling stage} is trivial: each node locally flips a biased coin,
so that the node enters $S$ with probability $p$ ($p$ is a parameter
to be fixed later).  This step is completely local, and by its end,
each node knows whether it is
a member of $S$ or not.

The \textbf{exploration stage} is the heart of our algorithm. To facilitate it,
we first construct a spanning tree for each connected components of $G[S]$
(Step \ref{st-B-tree} of the exploration stage). This construction is
implemented by constructing a BFS spanning tree of each connected component
$S_i$, rooted at the node with the smallest ID in $S_i$. This is a standard
distributed procedure (see, e.g., \cite{Peleg:book}), but here only the nodes
in $S$ take part, and all other nodes are non-existent for the purpose of this
protocol.

In Step \ref{st-B-ids} of the exploration stage, all nodes send their IDs to
the root. Once the root has all IDs, it sends them back down the tree.

In Step \ref{st-B-neigh} of the exploration stage, each node in $S_i$ sends the
identity of all nodes in $S_i$ to all
its neighbors. 
In addition, we effectively add to
each spanning tree all adjacent nodes. This is important so that we avoid
over-counting later. Note that a node of $S$ is member of a
single tree (the
tree of its connected component), but a node in $V\setminus S$ may have more
than one parent pointer: it has exactly one pointer for each component it is
adjacent to.

Step \ref{st-B-count} of the exploration stage determines for each node its
membership in $T_\eps(X)$ for each subset $X$ of each connected component.
Consider a node $u\in \Gamma(S_i)$. After Step \ref{st-B-neigh}, $u$ knows the
IDs of all members of $S_i$, so it can locally enumerate all $2^{|S_i|}$
subsets $X\subseteq S_i$, and furthermore, $u$ can determine whether $u\in
K_{2\eps^2}(X)$ for each such subset $X$. Thus, each such node $u$ locally
computes $2^{|S_i|}$ bits: one for each possible subset $X\subseteq S_i$. We
assume that the coordinates of the resulting vector are ordered in a well known
way (say, lexicographically). These vectors are sent by each node
$u\in\Gamma(S_i)$ to all its neighbors, and in particular to its parent in
$S_i$. This is done by $u$  for each $S_i$ it is adjacent to.  Step
\ref{st-B-convg} is implemented using standard convergecast on the tree
spanning $S_i$: the vectors are summed coordinate-wise and sent up the tree, so
that when the information reaches the root of $S_i$, it knows the size of
$K_{2\eps^2}(X)$ for each $X\subseteq S_i$. Finally, using the size of
$K_{2\eps^2}(X)$, and  knowing which of its neighbors is in $K_{2\eps^2}(X)$,
each node $u$ can determine whether $u\in K_\eps(K_{2\eps^2}(X))$, and thus
decide whether it is in $T_\eps(X)$ for each of the possible subsets $X$.

\ignore{ We summarize the time complexity of Algorithm \ref{alg-compute} as
follows.
\begin{lemma} \label{lem-alg-B}
  Algorithm \ref{alg-compute} runs in $O(2^{|S|})$ rounds in
  the worst case, using messages of $O(\log n)$ bits.
\end{lemma}
\Proof As discussed above, Steps \ref{st-B-tree}--\ref{st-B-neigh} take at most
$O(|S|)$ rounds. Step \ref{st-B-det} requires no communication. Step
\ref{st-B-add} requires a node $u$ to send $0$ or $1$ for each subset of each
component of $S$ it is adjacent to. Since there may be at most $2^{|S|}$ such
subsets (over all components), this step takes at most $O(2^{|S|})$ rounds. In
Step \ref{st-B-convg}, each entry of the vector may be a number between $0$ and
$n$, and hence the total number of bits is a vector is at most $2^{|S|}\log n$;
using pipelining once again, we can therefore bound the number of rounds
required to execute Step \ref{st-B-convg} by $O(2^{|S|}+|S|)=O(2^{|S|})$.
Similarly for Steps \ref{st-B-bcast}--\ref{st-B-K2}. Step \ref{st-B-dec} is
local. \QED}

When the \textbf{decision stage} of Algorithm $\dnc$ starts, each connected
component $S_i$ of $G[S]$ has a ``candidate'' near-clique and we need to choose
the largest $T_\eps(X)$ over all $X$'s. The difficulty is that there may be
more than one set that qualifies as a near-clique, and these sets may overlap.
Just outputting the union of these sets may be wrong because in general, the
union of \ENCe\ need not be an \ENCe. The decision stage resolves this
difficulty by allowing each node to ``vote'' only for the largest subset it is
a member of. This vote is implemented by killing all other subsets using
`abort' messages, which is routed to the root of  the spanning tree constructed
in the exploration stage. This ensures that from each collection of overlapping
sets, the largest one survives. Some small node sets may also have non-$\bot$
output: they can be disqualified if a lower
bound on the size of the dense subgraph is known. 

\subsection{Wrappers}
To conclude the description of the algorithm, we explain how to
obtain a deterministic upper bound on the
running time, and how to decrease error probability.

\noindent$\bullet$
\emph{Bounding the running time.} As we argue in Section
\ref{ssec-comp},
the time complexity of the algorithm can be bounded with some constant
probability.
If  a deterministic  bound on the running time is desired, one can add a
counter at each node, and abort the algorithm
if the running time exceeds the specified time limit.

\noindent$\bullet$
\emph{Boosting the success probability.}
The way to decrease the failure probability is not simply running the
algorithm multiple times. Rather, only  the
sampling and exploration stages are run several times independently, and then
apply a single decision stage to select the output. More specifically,
say we want to achieve success probability of at least $1-q$ for some
given $q>0$. Let $\lambda\DEF\log_{1-r}q$. To get failure probability
at most $q$, we run $\lambda$ independent
versions of the sampling and exploration stages (in any interleaving
order). These $\lambda$ versions are run with a deterministic time
bound as explained above. When  all versions terminate, a single decision
stage is run, and in  Step \ref{st-C-abort} of the
decision stage,  nodes
consider candidates from all $\lambda$ versions, and choose (by
sending ``acknowledge'') only the
largest of these candidates. 
This boosting wrapper increases the running time by a factor
of $\lambda$: the sampling and exploration stages are run $\lambda$
times, and the decision stage is slower by a factor of
$\lambda$ due to congestion on the links.

\section{Analysis}
\label{sec-analysis}
In this section we sketch the analysis of Algorithm $\dnc$ presented in Section
\ref{sec:alg}. Full details (i.e.,  most proofs) are presented in the appendix.

\subsection{Complexity}
\label{ssec-comp}
We first state the time complexity in terms of the sample size, and
then bound the sample size.

\begin{lemma}\label{lem:roundcomp}
Let $S$ be the set of nodes sampled in the sampling stage of Algorithm $\dnc$.
Then the round complexity of the algorithm is at most $\OO{2^{\abs{S}}}$.
\end{lemma}

\begin{lemma}\label{lem:boundedsample}
  $\Pr[\abs{S} \le 2 p n]\ge 1-e^{-\frac{p n}{3}}$.
\end{lemma}

\subsection{Correctness}
\label{ssec-correctness}

In this section we prove that Algorithm $\dnc$ finds a large near-clique. We
note that while the algorithm appears similar to the $\rho$-clique algorithm in
\cite{GGR}, the analysis of Algorithm $\dnc$ is different. We need to account
for the fact that the input
 contains a near-clique (rather than a clique), and we need to
 establish certain
 locality properties to show feasibility of a distributed implementation.

For the remainder of this section, fix $G=(V,E)$, $\epsilon>0$, and $\delta
>0$.  Let $\abs{V}=n$. Assume that $D\subseteq V$ is an $\epsilon^3$-near
clique satisfying $\abs{D} \ge \delta n$. Recall that $G[S]$ denotes the
subgraph of $G$ induced by $S$. In addition, assume that $\epsilon < \inv{3}$
(larger values are meaningless, see parameters of Theorem \ref{thm-detailed}).
\Znote{``In addition, assume w.l.o.g.\ that $\epsilon < \inv{3}$ (otherwise,
the algorithm can be applied with $\eps'\gets\inv3$).'' This is not true. If
the graph has a \ENC{$\eps^3$} it may not have \ENC{$\eps'^3$}.}

Let $D'$ denote the set of nodes output by Algorithm $\dnc$. Clearly,
$D'=T_{\eps}(X)$ for some $X$. We first show that every $T_{\eps}(X)$ is
\ENC{$\frac{n}{t}\eps$} where $t=|T_{\eps}(X)|$. 
In  the decision stage, the algorithm  selects the
largest $T_{\eps}(X)$. In Lemma \ref{lem:xstar}, we prove our main
technical result, namely that with constant probability, there exists a
subset
$X^* \subseteq S_i$ with $\abs{T_{\eps}(X^*)} \ge
(1-(\eps))\abs{D}$.

\paragraph{All large $T_{\eps}(X)$ are near-cliques.} The following lemma
proves that any $T_{\eps}(X)$ is a near-clique with a parameter relating to its
size. 

%

\begin{lemma}
\label{lem:density}
Let $X \subseteq V$, and denote $t = \abs{T_{\epsilon}(X)}$. Then
$T_{\epsilon}(X)$ is \ENC{$\frac{n\eps}{t}$}.
\end{lemma}

\paragraph{Existence of a large $T_{\eps}(X)$.} We  prove the
existence of a connected set $X^*\subseteq S$ such that
$T_{\eps}(X^*)$ is large.

First, let $C$ denote the set of all
nodes in the \ENC{$\eps^3$} $D$ that are also adjacent to all but $\eps^2$
fraction of $D$. Formally:
$\displaystyle
C \DEF K_{\epsilon^2}(D)\cap D~\mbox{ where $D$ is \ENC{$\eps^3$}.}
$
We use the following simple property.
\begin{lemma}
\label{lem-C}
  $\abs{C} \ge (1-\eps)\abs{D}-{1\over\eps^2}$.
\end{lemma}

Second,
we structure
the probability space defined by the sampling stage of Algorithm $\dnc$ as
follows. In the algorithm, each node flips a coin with probability $p$
of getting ``heads'' (i.e., entering $S$). We view this as  a
two-stage process, where each node flips \emph{two} independent coins: ${\sf coin}_1$
with probability $p_1\DEF p/2$ of getting ``heads'' and ${\sf coin}_2$ with
probability $p_2\DEF\frac{p-p_1}{1-p_1} > p/2$ of getting ``heads.'' A node
enters $S$ iff at least one of its coins turned out to be ``heads.''
The idea is that
the net result of the process is that each node enters $S$ independently with
probability $p$, but this refinement allows us to define two subsets of $S$:
let $S^{(1)}$ be the set of nodes for which ${\sf coin}_1$ is heads, and let
$S^{(2)}$ be the set of nodes for which ${\sf coin}_2$ is heads.

Combining the notions, we define $X^*\DEF S^{(1)} \cap C$, i.e., $X^*$ is a
random variable representing the set of nodes from $C$ for which ${\sf coin}_1$
is heads. $X^*$ is effectively a sample of $C$ where each node is selected with
probability $p/2$. We have the following.
\begin{lemma}
\label{lem-single}
  $X^*$ resides within a single connected component of $G[S]$ with
  probability at least $1-e^{-\OM{\delta pn}}$.
\end{lemma}

We now arrive at our main lemma.
\begin{lemma}
\label{lem:xstar} With probability at least
$1-\inv{\eps^2\delta}e^{-\OM{\eps^4\delta\cdot pn}}$ over the selection of $S$,
there exists a connected component $S_i$ of $G[S]$ and a set $X^* \subseteq
S_i$ s.t.\ $\abs{T_{\eps}(X^*)} \ge (1-\frac{13}{2}\eps)\abs{D}-\eps^{-2}$.
\end{lemma}


\Proof
Let $X^*$ be defined as above.
It remains to show that $T_{\eps}(X^*)$ is large. Intuitively, $X^*$ is a
random sample of $C$, and since $C$ contains almost all of $D$, $X^*$ is also,
in a sense, a sample of $D$. Thus $K_{2\eps^2}(X^*)$ should be very close to
$K_{(\cdot)}(C)$, $K_{(\cdot)}(D)$ for appropriately selected $(\cdot)$. This
would complete the proof since $T_{\eps}(C)$ contains almost all of $C$ which,
in turn, contains almost all of $D$. Formally, we say that $X^*$ is
\emph{representative} if the following hold.
\begin{denseenumerate}
\item\label{item:KXlower} $\abs{K_{\epsilon^2}(D) \setminus
    K_{2\epsilon^2}(X^*)} < \epsilon\abs{C}$.
\item\label{item:KXupper} $\abs{K_{2\epsilon^2}(X^*) \setminus
    K_{3\epsilon^2}(C)} <\epsilon^2\abs{C}$.
\end{denseenumerate}
That is, if $K_{2\eps^2}(X^*)$ is almost fully contained in $K_{\eps^2}(D)$ and
almost fully contains $K_{3\eps^2}(C)$.

To complete the proof, we use two claims presented below. Claim
\ref{claim:ifrep} shows that if $X^*$ is representative, then $\abs{C \setminus
T_{\epsilon}(X^*)} \le \frac{11}{2}\eps\cdot\abs{C}$. Claim
\ref{claim:indeedrep} shows  that $X^*$ is representative with probability
$1-\inv{\eps^2\delta}e^{-\OM{\eps^4\delta pn}}$. Given these claims, the proof
is completed as follows. By Lemma \ref{lem-single} and the claims, we have that
$1-(e^{-\Omega(\delta p n)}-\inv{\eps^2\delta}e^{-\OM{\eps^4\delta
    pn}})$, we have that $X^*$ resides in a connected component of
$G[S]$, and, using also Lemma \ref{lem-C} the proof is complete, because
\[
\abs{T_{\eps}(X^*)} ~\ge~ \left(1-\frac{11}{2}\eps\right)\abs{C} ~\ge~
\left(1-\frac{11}{2}\eps\right)\left((1-\eps)\abs{D}-{1\over\eps^2}\right)
~\ge~\left(1-\frac{13}{2}\eps\right)\abs{D}-{1\over\eps^2}~.\quad\QED
\]

\begin{claim}\label{claim:ifrep}
If $X^*$ is representative, then $\abs{C \setminus T_{\epsilon}(X^*)} \le
\frac{11}{2}\eps\cdot\abs{C}$
\end{claim}

\begin{claim}\label{claim:indeedrep}
$\Pr\left[X^* \mbox{ is representative}\right] \ge
1-\inv{\epsilon^2\delta}\cdot e^{-\OM{\epsilon^4\delta pn}}$.
\end{claim}

\subsection{Summary}
\label{ssec-summary}
We summarize with the following theorem, which is the detailed version
of  Theorem \ref{thm:algorithm-hl} (in Theorem \ref{thm:algorithm-hl},
we set $p = \inv{n} \cdot
O\left(\frac{\log(\inv{\epsilon\delta})}{\epsilon^4\delta}\right)$).

\begin{theorem}
\label{thm-detailed}
Let $G=(V,E)$, $\abs{V}=n$. Let $D \subseteq V$ be an
$\epsilon^3$-near clique in $G$ of size $\abs{D} \ge \delta n$. Then
with probability at least $1-\frac{1}{\epsilon^2\delta}\cdot
e^{-\OM{\epsilon^4\delta \cdot p n}}$, Algorithm $\dnc$, running on
$G$ with parameters $\epsilon, p$, finds, in $\OO{2^{2pn}}$
  communication rounds, a subgraph $D'$ such that
\begin{denselist}
\item\label{i:enc}
  $D'$ is
  \ENC{$\left(\inv{(1-\frac{13}{2}\epsilon)}\cdot\frac{\epsilon}{\delta}\right)$}.%
\footnote{For
small enough $\epsilon$, say $\epsilon<\inv{13}$, this is at most
$2\frac{\epsilon}{\delta}$.}
\item\label{i:big}
  $\abs{D'} \ge (1-\frac{13}{2}\epsilon)\abs{D}-\eps^{-2}$.
\end{denselist}
 \end{theorem}

\Proof By Lemmas
\ref{lem:roundcomp} and
\ref{lem:boundedsample}, the probability that the round complexity
exceeds $2^{\OO{2pn}}$ is bounded
by $e^{-\frac{pn}{3}}$.
By Lemma \ref{lem:density}, whenever assertion (\ref{i:big})
holds, assertion (\ref{i:enc}) holds as well.
Assertion (\ref{i:big}) holds by Lemma \ref{lem:xstar} with probability
at least $1-\inv{\eps^2\delta}e^{-\OM{\eps^4\delta\cdot pn}}$.
The theorem follows from the union bound.
\QED

\Znote{Removed wrong remark and moved part of discussion here.} \ignore{We
remark that the local computation done by each node in the algorithm requires
$\OO{2^{\abs{S}}}$ operations on $\OO{\log n}$ bit numbers. For our purposes,
$|S|\le O(\log\log n)$.}

It may also be interesting to analyze the computational complexity of the
vertices running the algorithm. A simple analysis shows that except for step
\ref{st-B-dec} of the exploration stage, the operation for each node can be
implemented in $\poly(\abs{S})$ computational steps (on $\log n$ bit numbers)
per communication round. In step \ref{st-B-dec}, however, the nodes need to
``inspect'' all their neighbors in order to determine whether they reside in
$T_{\eps}(X)$. It is possible to reduce the complexity in this case by
selecting a sample of the neighbors and estimating, rather than determining,
membership in $T_{\eps}(X)$. Thus, the computational complexity can be reduced
to $\poly(\abs{S})$ computational steps per round (for our purposes, $|S|\le
O(\log\log n)$). The analysis of this modification is omitted.

\section{Discussion}
\label{sec:discussion}
\paragraph{On the impossibility of finding a globally maximal \ENCe.} Our
algorithm (when successful) finds a disjoint collection of
near-cliques such that at least one of them is large. We note that  it
is impossible for a distributed sub-diameter time algorithm
to output just one (say, the largest) clique.
To see that, consider a graph containing an $n/2$-vertex clique $A$ and an
$n/4$-vertex clique $B$, connected by an $n/4$-long path $P$. The largest
near-clique in this case is obviously $A$, and the vertices of $B$ should
output $\bot$. However, if we delete all edges in $A$, the largest near-clique
becomes $B$, i.e., its output must be non-$\bot$. Since no node in $B$ can
distinguish between the two scenarios in less than  $|P|=n/4$ communication
rounds, impossibility follows.


\paragraph{Deriving distributed algorithms from property testers.} Our approach
may raise hopes that other property testers, at least in
the dense graph model,\footnote{We note that the dense-graph model is, in many
cases, inadequate for modeling communication networks as such graphs are often
sparse (and thus a solution for an $\eps$-close graphs is either trivial or
uninteresting).} can be adapted into the distributing setting. Goldreich and
Trevisan \cite{GT03} prove that any property tester in the dense graph model
has a canonical form where the first stage is selecting a uniform sample of
appropriate size from the graph and the second is testing the graph induced by
the sample for some (possibly other) property. Thus, the following scheme may
seem likely to be useful:\\
1. Select a uniform sample by having all nodes flip a biased coin.\\
2. Find the graph induced between sampled nodes. This graph has very
small (possibly constant) size. \\
3. Use some (possibly inefficient) distributed algorithm to test it for
the required property.
\\
In the distributed setting, however, sometime even testing a property for a
very small graph would be impossible due to connectivity issues. As
demonstrated above, there exist properties that are testable in the centralized
setting and do not admit an efficient round-complexity distributed algorithm.
The general method above, therefore, can only be applied in a ``black-box''
manner for some testers.

Specifically, the $\rho$-clique tester presented in \cite{GGR} does not comply
with the above requirements (specifically, as we mentioned, the $\rho$-clique
problem is unsolvable in small round-complexity). It can, however, be converted
into a near-clique finder, in the sense defined in this work, using similar
ideas and with worse parameters.

\newpage
\pagenumbering{roman}
{

}
\newpage
\section*{APPENDIX: Additional Proofs}
\ProofOf{Lemma}{lem:roundcomp}
The sampling stage requires no communication. Consider now the
exploration stage.
The BFS tree construction of Step \ref{st-B-tree} uses messages of $O(\log n)$
bits (each message contains an ID and a distance counter), and its running time
is proportional to the diameter of the component, which is trivially bounded by
$|S|$.
The number of rounds to execute Step \ref{st-B-ids} is proportional to
the number of IDs plus the height of the tree, due to the pipelining
of messages: the number of hops each ID needs to travel is at most
twice the tree height, and a message needs to wait at most once for
each other ID. It follows that the total time required for this step
in $O(|S|)$ rounds.
Step \ref{st-B-neigh} takes at most $\max_i\Set{|S_i|}\le|S|$ rounds.
Step \ref{st-B-det} requires no communication.
Step \ref{st-B-add} requires a node $u$ to send $0$ or $1$ for each subset of
each component of $S$ it is adjacent to. Since there may be at most $2^{|S|}$
such subsets (over all components), this step takes at most $O(2^{|S|})$
rounds. In Step \ref{st-B-convg}, each entry of the vector may be a number
between $0$ and $n$, and hence the total number of bits in a vector is at most
$2^{|S|}\log n$; using pipelining once again, we can therefore bound the number
of rounds required to execute Step \ref{st-B-convg} by
$O(2^{|S|}+|S|)=O(2^{|S|})$.  Similarly for Steps
\ref{st-B-bcast}--\ref{st-B-K2}. Step \ref{st-B-dec} is local; In the decision
stage, Step \ref{st-C-size} takes, again, at most $O(2^{|S|})$ rounds. The
remaining steps take at most $O(\max_i \abs{S_i}) \le O(\abs{S})$ rounds. Thus
the total round complexity of Algorithm $\dnc$ is at most $\OO{2^{\abs{S}}}$.
\QED

\ProofOf{Lemma}{lem:boundedsample}
Follows from the Chernoff  Bound, since in the sampling stage, each of the $n$
nodes join $S$ independently with probability $p$.
\QED

\ProofOf{Lemma}{lem:density}
By counting.
Recall that each undirected
edge is viewed and counted as two anti-symmetrical directed edges.
Define $Y = K_{2\epsilon^2}(X)$.
Consider a node  $v \in T_{\epsilon}(X)$. By definition of $T_\eps(X)$,
  $v \in K_{\epsilon}(Y)$, i.e.,
  \begin{equation}
    \label{eq:Y}
    \abs{\Gamma(v) \cap Y} ~\ge~ (1-\epsilon)\abs{Y}~.
  \end{equation}
Since $T_{\epsilon}(X) \subseteq Y$, we have $\abs{\Gamma(v) \cap
T_{\epsilon}(X)} ~\ge~ \abs{\Gamma(v) \cap Y} ~\ge~ t - \epsilon\abs{Y}$ by
\Eqr{eq:Y}. Since $|Y|\le n$, we can conclude that $\abs{\Gamma(v) \cap
T_{\epsilon}(X)} \ge (1-\frac{n}{t}\epsilon)t$. It follows that the total
number of (directed) edges in  $T_{\epsilon}(X)$ is at least
$(1-\frac{n}{t}\epsilon)t(t-1)$, as required. \QED
\QED

\ProofOf{Lemma}{lem-C}
Denote
$c\DEF|C|$ and $d\DEF|D|$. Since $D$ is an
\ENC{$(1-\eps^3)$} in $G$,  we have that
\begin{equation}
  \label{eq:d}
|E\cap(D\times D)|
~\ge~(1-\eps^3)d(d-1)
~\ge~(1-\eps^3)d^2-d~.
\end{equation}
By definition of $C$, if $v\in D\setminus C$, then
\begin{equation}
  \label{eq:d+1}
  |E\cap (\Set{v}\times D)|<(1-\eps^2)d~.
\end{equation}
Now, if we assume that $c<(1-\eps-{1\over\eps^2 d})d$, we arrive at a
contradiction to \Eqr{eq:d}, since
\begin{eqnarray*}
|E\cap(D\times D)|
&=&|E\cap(C\times D)|+|E\cap((D\setminus C)\times D)|\\
&\le&c\cdot d+(d-c)(1-\epsilon^2)d
~~~~~~~~~~~~~~~~~~~~~~~~~~~~~~~\mbox{by \Eqr{eq:d+1}}\\
& =&(1-\epsilon^2)\cdot d^2 + \epsilon^2\cdot c\cdot d\\
& <& (1-\epsilon^2)\cdot d^2
+\epsilon^2\left((1-\epsilon)d-{1\over \eps^2}\right)d
~~~~~~~~~~~~~~~\mbox{if $c<(1-\eps)d-{1\over \eps^2}$}\\
&\le&(1-\epsilon^3)d^2-d~.
\end{eqnarray*}
\QED

\ProofOf{Lemma}{lem-single}
We show that a stronger property holds with that probability:
namely, that the
distance in $S$ between any two nodes of $X^*$ is at most $2$. By definition,
$X^* \subseteq C$, i.e.,  $X^* \subseteq K_{\epsilon^2}(D)$. It follows from
the pigeonhole principle that every two nodes $u, v \in X^*$ have at least
$(1-2\epsilon^2)\abs{D}$ common neighbors. The probability that none of these
common neighbors is in $S^{(2)}$ (i.e., that none of them has outcome heads for
${\sf coin}_2$) is therefore at most $(1-p_2)^{(1-2\epsilon^2)\abs{D}} \le
e^{-(1-2\epsilon^2)p_2\abs{D}} \le e^{-\frac{7}{18}p\abs{D}}$ for $\eps\le
1/3$, and because $p_2 > p/2$ by definition. We now apply the union bound to
obtain that $\Pr\left[\mbox{diameter}(X^*)>2\right] \le\abs{X^*}\cdot
e^{-\frac{7}{18}p\abs{D}}$. Since $X^*$ is a random sample of $C$, it follows
that $\bbE[\abs{X^*}] = p_2 \cdot \abs{C} \le p\delta n$. Using a
Chernoff bound, we obtain that $\Pr[\abs{X^*} > 2
p \delta n] \le e^{-\OM{\delta pn}}$. Therefore, by the union bound
\begin{eqnarray*}
\Pr\left[\mbox{diameter}(X^*)>2\right]
& \le & \Pr\left[\mbox{diameter}(X^*)>2~|~ \abs{X^*} \le 2p \delta n\right]
+ \Pr\left[\abs{X^*} > 2p \delta n\right]\\
&\le & 2\delta pn \cdot e^{-\OM{\delta pn}}+e^{-\OM{\delta pn}} \le
e^{-\OM{\delta pn}}~.\qquad\qquad\qquad\qquad\qquad\QED
\end{eqnarray*}

\ProofOf{Claim}{claim:ifrep}
By definition,
\begin{equation}\label{eq:t0}
\abs{C \setminus T_{\epsilon}(X^*)} ~\le~ \abs{C \setminus K_{2\epsilon^2}(X^*)} +
\abs{C \setminus K_{\epsilon}(K_{2\epsilon^2}(X^*))}~.
\end{equation}
We bound  each term in \Eqr{eq:t0} in turn.

First, note that $\abs{K_{\epsilon^2}(D)\setminus  K_{2\epsilon^2}(X^*)} <
\epsilon\abs{C}$, because $X^*$ is representative. It follows that
\begin{equation}
  \label{eq:t1}
  \abs{C \setminus K_{2\epsilon^2}(X^*)} < \epsilon\abs{C}~,
\end{equation}
because $C \subseteq K_{\epsilon^2}(D)$. Note that \Eqr{eq:t1} also implies for
$\eps\le1/3$ that
\begin{equation}
  \label{eq:t11}
  \abs{K_{2\epsilon^2}(X^*)}~ \ge~ (1-\epsilon)\abs{C}
~\ge~ \frac{2\abs{C}}{3}~.
\end{equation}

We now turn to the second term of \Eqr{eq:t0}.
 $X^*$ is
representative, and therefore $\abs{K_{2\epsilon^2}(X^*) \setminus
K_{3\epsilon^2}(C)} <\epsilon^2\abs{C}$, i.e., all but $\epsilon^2\abs{C}$
vertices of $K_{2\epsilon^2}(X^*)$ are neighbors of at least
$(1-3\epsilon^2)|C|$ nodes of $C$. Let $Y = C \setminus
K_{\epsilon}(K_{2\epsilon^2}(X^*))$, $y = \abs{Y}$, and $z =
\abs{K_{2\epsilon^2}(X^*)}$. Counting the number of edges between $C$ and
$K_{2\epsilon^2}(X^*)$ we conclude that \( y\cdot(1-\epsilon)z +
(\abs{C}-y)\cdot z \ge \left(z-\epsilon^2\abs{C}\right)(1-3\epsilon^2)\abs{C}
\), and plugging in \Eqr{eq:t11} we obtain
\[
y\cdot(1-\epsilon)z + (\abs{C}-y)\cdot z \ge z\cdot
(1-\frac{3\epsilon^2}{2})(1-3\epsilon^2)\abs{C} \ge
(1-\frac{9\epsilon^2}{2})\cdot z \abs{C}~.
\]
Rearranging, we have $y \le \frac{9\epsilon}{2}\cdot\abs{C}$, and the claim
follows.
\QED

\ProofOf{Claim}{claim:indeedrep}
Since $\bbE[\abs{X^*\!}]=p_1\abs{C}$, and since membership in $X^*$ is
determined independently for each node, we can apply  the Chernoff Bound to
obtain that
\[
\Pr\left[\Big|\abs{X^*\!}-\bbE\big[\abs{X^*\!}\big]\Big|
> \frac{\epsilon^2}{4}\bbE\big[\abs{X^*\!}\big]\right]
~<~
2 \exp\left({-\inv{3}\left(\frac{\epsilon^2}{4}\right)^2
   \bbE\big[\abs{X^*\!}\big]}\right)
~\le~
 2\exp\left(-{\epsilon^4 p_1\abs{C}\over48}\right).
\]
Assume that $\Big|{\abs{X^*}-\bbE\big[\abs{X^*\!}\big]}\Big| \le
\frac{\epsilon^2}{4}\bbE\big[\abs{X^*\!}\big]$, and let us consider the
definition of a representative set.

For  item \ref{item:KXlower}\ignore{ of Definition \ref{def-rep}},  let $v \in
K_{\epsilon^2}(D)$. Then
$\abs{\Gamma(v) \cap C} \ge \abs{C}-\epsilon^2\abs{D} \ge
\abs{C}-\frac{\epsilon^2}{1-\epsilon}\abs{C} \ge
(1-\frac{3}{2}\epsilon^2)\abs{C}$. Since
 $\Gamma(v) \cap X^*$ is a random
sample of $\Gamma(v) \cap C$, where each member is chosen  with probability
$p_1$, we have that $\bbE\left[\abs{\Gamma(v) \cap X^*}\right] = p_1 \cdot
\abs{\Gamma(v) \cap C} \ge
(1-\frac{3}{2}\epsilon^2)p_1\abs{C}=(1-\frac{3}{2}\epsilon^2)\bbE[\abs{X^*}]$.
Denote $Y_v = \abs{\Gamma(v) \cap X^*}$. Then
\begin{eqnarray*}
\Pr[v \not\in K_{2\epsilon^2}(X^*)]
& = & \Pr[Y_v < (1 - 2\epsilon^2)\abs{X^*\!}]
~\le~
\Pr\Big[Y_v < (1 - 2\epsilon^2)(1+\frac{\epsilon^2}{4})\bbE[\abs{X^*\!}]\Big]\\
& \le &
\Pr\Big[Y_v -\bbE[Y_v] < (1 -2\epsilon^2)(1+\frac{\epsilon^2}{4})\bbE[\abs{X^*\!}]-(1-\epsilon^2)\bbE[\abs{X^*\!}]\Big]\\
& \le & \Pr[Y_v -\bbE[Y_v] < -\frac{\epsilon^2}{4}\bbE[\abs{X^*\!}]]\\
& < &
\exp\left(-\inv{2}\left(\frac{\frac{\epsilon^2}{4}\bbE[\abs{X^*\!}]}{\bbE[Y_v]}\right)^2\bbE[Y_v]\right)\\
& \le & \exp\left(-{\epsilon^4\over32}\bbE[\abs{X^*\!}]\right)
~\le~
\exp\left(-{\epsilon^4\over32}\cdot p_1\abs{C}\right),
\end{eqnarray*}
and therefore $\bbE[\abs{K_{\epsilon^2}(C) \setminus K_{2\epsilon^2}(X^*)}] <
\exp\left(-{\epsilon^4\over32} \cdot p_1\abs{C}\right) \cdot n$.  Using
Markov's Inequality we obtain
\[
\Pr\Big[\abs{K_{\epsilon^2}(C) \setminus K_{2\epsilon^2}(X^*)} \ge
\epsilon\abs{C}\Big]
~ \le~
\frac{n}{\epsilon\abs{C}}\cdot
e^{-{\epsilon^4} \cdot p_1\abs{C}/32}.
\]

A similar argument applies to item \ref{item:KXupper}\ignore{ of Definition
\ref{def-rep}}. Consider a node $v \not\in
K_{3\epsilon^2}(C)$
. Denote $Y_v = \abs{\Gamma(v) \cap X^*}$. Then $\bbE[Y_v] <
(1-3\epsilon^2)\bbE[\abs{X^*\!}]$, and therefore 
\begin{eqnarray*}
\Pr[v \in K_{2\epsilon^2}(X^*)]
& = & \Pr[Y_v \ge (1 - 2\epsilon^2)\abs{X^*\!}]
~\le~ \Pr\Big[Y_v \ge (1 - 2\epsilon^2)
   (1-\frac{\epsilon^2}{4})\bbE[\abs{X^*\!}]\Big]\\
& \le &
\Pr\Big[Y_v -\bbE[Y_v]
     \ge(1 -2\epsilon^2)(1-\frac{\epsilon^2}{4})\bbE[\abs{X^*\!}]
          -(1-3\epsilon^2)\bbE[\abs{X^*\!}]\Big]\\
& \le &
\Pr\Big[Y_v -\bbE[Y_v] \ge \frac{3\epsilon^2}{4}\bbE[\abs{X^*\!}]\Big]\\
& < &
\exp\left(-\inv{3}\left(\frac{\frac{3\epsilon^2}{4}\bbE[\abs{X^*\!}]}
 {\bbE[Y_v]}\right)^2\bbE[Y_v]\right)\\
& \le &
\exp\left(-\frac{3\epsilon^4}{16}\cdot\bbE[\abs{X^*\!}]\right)
~\le~
\exp\left(-\frac{3\epsilon^4}{16} \cdot p_1\abs{C}\right),
\end{eqnarray*}
i.e., $\bbE[\abs{K_{2\epsilon^2}(X^*) \setminus K_{3\epsilon^2}(C)}] <
\exp\left(-\frac{3}{16}\epsilon^4 \cdot p_1\abs{C}\right) \cdot n$, which
implies, as above, that
\[
\Pr\Big[\abs{K_{2\epsilon^2}(X^*) \setminus K_{3\epsilon^2}(C)} \ge \epsilon^2
\abs{C}\Big]
~\le~ \frac{n}{\epsilon^2 \abs{C}}\cdot e^{-3\epsilon^4
\cdot p_1\abs{C}/16}.
\]
Finally, we apply the Union Bound to that $X^*$ is representative with
probability at least
\[
1-\left(2e^{-\epsilon^4 p_1\abs{C}/48}+ \frac{n}{\epsilon\abs{C}}\cdot
e^{-\epsilon^4 \cdot p_1\abs{C}/32}+ \frac{n}{\epsilon^2 \abs{C}}\cdot
e^{-{3}\epsilon^4 \cdot p_1\abs{C}/16}\right) ~\ge~
1-\frac{n}{\epsilon^2\abs{C}}\cdot e^{-\OM{\epsilon^4 \cdot
p\abs{C}}}~.\quad\QED
\]

\end{document}